\documentclass[reprint,superscriptaddress,aps,prb,twocolumn]{revtex4-2}
\usepackage{setspace}
\usepackage{amsmath}
\usepackage{graphicx}
\usepackage{subfig}
\usepackage{ragged2e}
\usepackage{amsfonts}
\usepackage{amssymb}
\usepackage{epstopdf} 
\usepackage{xcolor}
\usepackage{verbatim}
\usepackage{soul}
\usepackage{comment}
\usepackage{placeins}
\usepackage{textcomp}
\usepackage{textgreek}
\usepackage{bm}
\usepackage{hyperref}

\DeclareGraphicsExtensions{.pdf,.eps,.png,.jpg,.mps} 

\usepackage{float}

\begin{document}

\title{Scalable 3D silicon nitride photonic interposer for high-density optical interconnects}

\author{
Yu Xia\textsuperscript{$\ast$},
Yuhao Huang\textsuperscript{$\ast$},
Yuemin Li,
Jie Wang,
Yunqi Fu,
Yaoran Huang,
Hongjie Liang,
Hao Fang,
Zheng Li,
Mingfei Liu,
Yitian Tong,
Di Yu,
and Chao Xiang\textsuperscript{$\dagger$}
\\
\vspace{0.5em}
$^1$Department of Electrical and Computer Engineering and State Key Laboratory of Optical Quantum Materials, The University of Hong Kong, Hong Kong, China\\
$\ast$These authors contributed equally to this work\\
$^\dagger$Corresponding author: cxiang@eee.hku.hk}

\begin{abstract}
Modern computing workloads demand energy-efficient, high-bandwidth interconnects, motivating photonic interposers as an alternative to electrical links. Here we demonstrate a compact 3D silicon nitride (SiN) photonic interposer prototype comprising two routing layers, with the 3D routing scheme optimized by a global optimization algorithm. The 3D interposer realizes a fully connected 12-node optical network that reduces the total number of intralayer crossings from 495 for all-planar routing to merely 150 (69.7\% reduction), below the theoretical lower bound of 153 for all-planar interconnects. Comparing the two schemes, our 3D design achieves a 45.8\% reduction experimentally in the average loss per waveguide. The proposed 3D routing architecture also features inherent symmetry and is scalable to higher node counts, flexible node placements, additional routing layers, and other operating wavelengths, enabling denser, lower-loss photonic interposers for next-generation scale-up and high-performance computing (HPC) systems.
\end{abstract}

\maketitle

Modern artificial intelligence (AI) and HPC workloads demand unprecedented data movement between compute units. In scale-up compute clusters, the interconnects between these units are increasingly becoming the bottleneck for system power consumption~\cite{gholamiAIMemoryWall2024b}.
Co-packaged optics (CPO) shortens the lossy copper path by placing optical engines adjacent to the application-specific integrated circuits (ASICs) and provides optical interconnects between these nodes~\cite{xiangBuilding3DIntegrated2024b}. 
While CPO eliminates long electrical paths, the growing needs for bandwidth are hindered by the practical difficulty and aggregated cost when a large number of fibers are needed. Moreover, the constraints from the chip shoreline density will ultimately limit the interconnect capacity and as a result, the system efficiency.

\begin{figure*}[!t]
  \centering
  \includegraphics[width=0.9\linewidth]{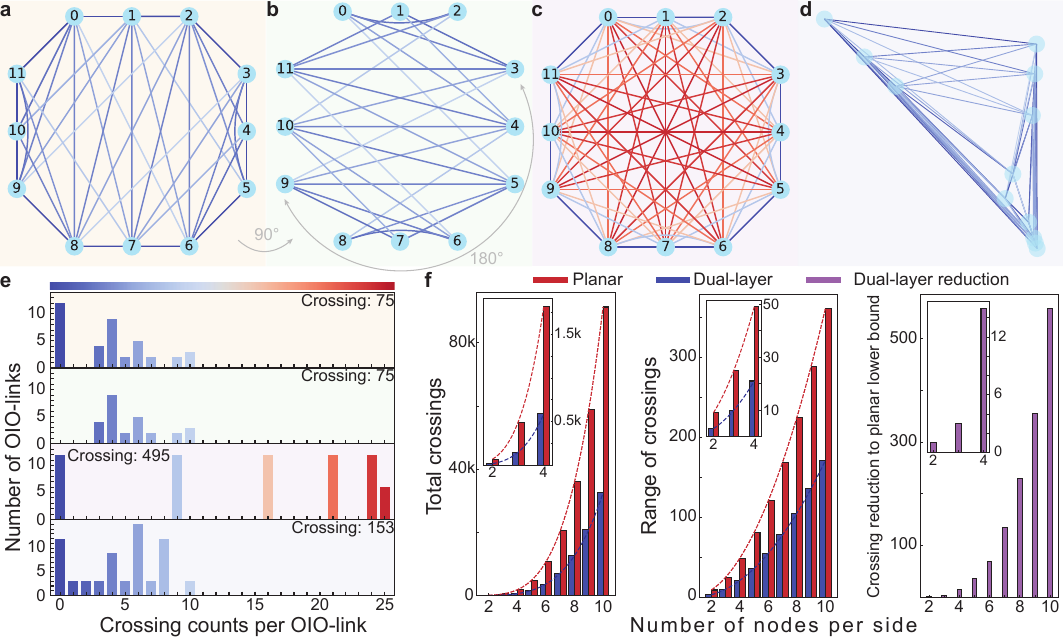}
  \caption{ Configuration for (a). Layer 0 and (b). Layer 1 of the optimized 3D routing, (c). all-planar routing and (d). all-planar routing with minimal rectilinear crossings~\cite{aichholzerOngoingProjectImprove2020}.
  (e). Histograms comparing crossing distributions for (a)–(d) from top to bottom, with color bar representing the crossing count.
  (f). Predicted crossing statistics versus nodes per side.}
  \label{fig_scheme}
\end{figure*}

In comparison, photonic interposers can route optical signals between optical I/Os (OIOs) in a highly-scalable manner that is insensitive to the increase of interconnect node counts. So far, several material platforms including polymer and glass waveguides have been explored. Polymer waveguides can achieve $\mathord{\sim}0.4\,\mathrm{dB/cm}$ propagation loss~\cite{vanaschLowlossIntegrationHighdensity2025} but the attenuation level may impose a constraint on the loss budget, given the potential requirement for link lengths on the order of tens of centimeters~\cite{Tauke-Pedretti:25}.
Glass waveguides, conversely, offer exceptionally low loss down to $\mathord{\sim}0.034\,\mathrm{dB/cm}$~\cite{brusbergUltraLowLossIonExchange2024a} but their large bending radii ($\mathord{\sim}3\,\mathrm{cm}$) restrict routing flexibility and density. Another challenge arising from all-planar routing is the rapid escalation of the unavoidable intralayer crossings. For instance, considering a fully connected $k=12$ nodes network with nodes placed uniformly around a square area, the total number of OIO-links is $\smash[b]{N_\text{link} = \binom{k}{2}}$ and the number of intralayer crossings on the $m$-th OIO-link, which connects node $0$ and node $m+1$, is $\smash[b]{C_m = m\,(k-2-m)}$, where $\smash[b]{0 \le m \le k-2}$. By symmetry, the same crossing profile applies to any node chosen as the reference. Consequently, the total number of crossings for all links is $\smash[b]{C_{\mathrm{tot}} = \binom{k}{4}}$.
These relationships imply that the maximum crossings per link scales as $\smash[b]{\propto k^2}$ and the total crossing count scales as $\smash[b]{\propto k^4}$ [Fig.~\ref{fig_scheme}(f)] for all-planar routing.
Because each OIO-link may be implemented by a bidirectional waveguide pair, one OIO-link-level crossing can correspond to four physical intralayer crossings, making the loss budget more constrained.
For clarity, the crossing throughout this letter refers to the OIO-link-level intralayer crossing unless stated otherwise.

To overcome these limitations, here we propose a 3D dual-layer SiN interposer prototype integrating 12 fully interconnected OIO-nodes. SiN is an attractive material for photonic integrated circuits due to its superior optical properties, such as wide transparency range, low thermo-optic coefficient and CMOS-compatible fabrication~\cite{xiangSiliconNitridePassive2022a}. Extending this platform into a multilayer 3D structure increases integration density~\cite{fannNovelParallelDigital2024} and routing flexibility by providing additional degrees of freedom in the vertical direction. Importantly, by routing waveguides across different layers, substantial high-loss intralayer crossings are converted to low-loss interlayer crossings ($\sim$10$\times$ - 1000$\times$ lower per crossing in~\cite{sacherMonolithicallyIntegratedMultilayer2018a}, and is 100$\times$ lower in this work). If incorporating efficient interlayer tapers, the reduced crossing losses directly deliver a significant reduction in overall link losses. Such improvement can be even more outstanding for systems with more nodes. The proposed interposer leverages the excellent passive properties of SiN and a novel 3D routing architecture, enabling low-loss, low-crosstalk, high-density and highly-scalable waveguide interconnects. While our approach readily scales to much higher node counts, the experimental prototype starts from a representative case of 12 nodes. 

In our implementation, fiber-to-chip coupling is conducted via edge couplers on Layer~0 (bottom layer) and interlayer tapers are introduced immediately after these edge couplers whenever routing on Layer~1 (top layer) is required. The 3D routing strategy is optimized based on the generalized simulated annealing (GSA) algorithm~\cite{tsallisGeneralizedSimulatedAnnealing1996}, which is a variant of the simulated annealing (SA) algorithm~\cite{kirkpatrickOptimizationSimulatedAnnealing1983}. The SA mimics the physical process in which a material anneals from an initial high-temperature, high-energy state to a low-temperature, low-energy state, while the GSA generalizes the SA for faster convergence and better exploration of the state space~\cite{xiangGeneralizedSimulatedAnnealing2013}. Here, the state space consists of a layer vector conveying layer assignments for all OIO-links, and the system energy is given by the average on-chip interconnect loss (ICL) over all OIO-links, which serves as the objective function to be minimized. In the following, the on-chip ICL indicates that the edge coupler and the external path losses are excluded. The optimization process is conducted through the dual annealing routine~\cite{virtanenSciPy10Fundamental2020}, which provides a robust implementation of the GSA. Comprehensive details of the optimization process are summarized in Supplement~1.

\begin{figure*}[!t]
  \centering
  \includegraphics[width=0.9\linewidth]{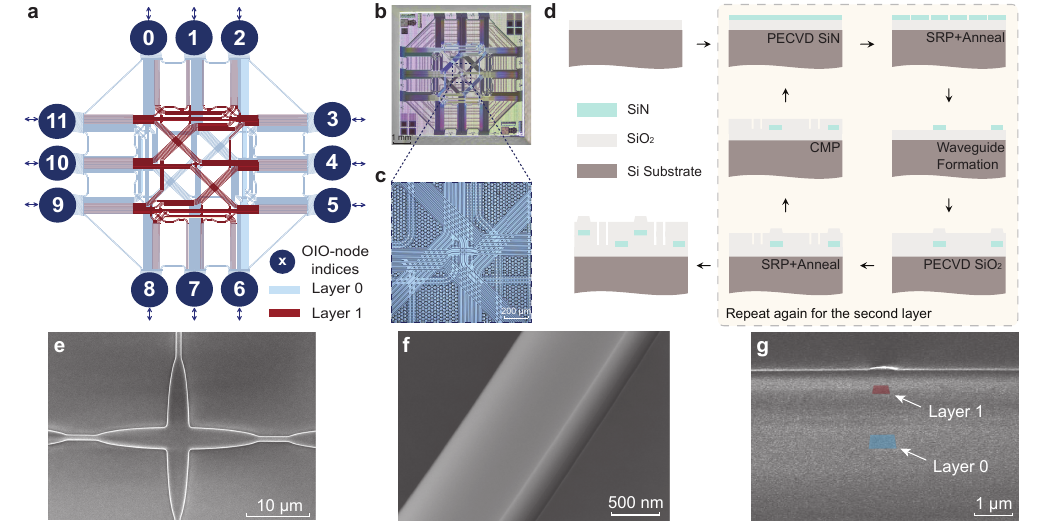}
  \caption{
  (a). Layout of the optimized 3D routing scheme for the dual-layer SiN photonic interposer containing 12 nodes. Layer~0 waveguides are colored blue, and Layer~1 waveguides are colored red.
  (b). Picture of the fabricated chip and (c). zoomed-in view of the central crossing region.
  (d). Schematic of fabrication process flow.
  SEM images of the key structures including (e). intralayer crossing, (f). SiN waveguide and (g). cross-sectional view of the interlayer taper.
  }
  \label{fig_dieshot}
\end{figure*}
The GSA-optimized 3D routing [Fig.~\ref{fig_scheme}(a)-(b)] drastically reduces the total number of intralayer crossings for a 12-node optical network from 495 in the conventional all-planar implementation [Fig.~\ref{fig_scheme}(c) and (e)] to merely 150, corresponding to a 69.7\% reduction and notably breaking the theoretical lower bound of 153 for all-planar interconnects~\cite{aichholzerOngoingProjectImprove2020} as shown in Fig.~\ref{fig_scheme}(c)-(e).
This trend persists as node count grows [Fig.~\ref{fig_scheme}(f), left].
Specifically, letting $n$ be the number of nodes per side ($k=4n$), the polynomial fits [dashed lines in Fig.~\ref{fig_scheme}(f)] reveal that the total number of crossings of the optimized dual-layer routing follows $\smash[b]{C'_{\mathrm{tot}}=n(n-1)(2n-1)^2}$ while the range of crossings scales as $\smash[b]{C'_{\mathrm{rng}}=(n-1)(2n-1)}$. Compared to the baseline of all-planar interconnects where $\smash[b]{C_{\mathrm{tot}} = \binom{4n}{4}}$ and $\smash[b]{C_{\mathrm{rng}}=(2n-1)^2}$, 
our approach is able to reduce the total crossing count and crossing range to $\smash[b]{[\frac{9}{35}, \frac{3}{8})}$ and $\smash[b]{[\frac{1}{3}, \frac{1}{2})}$, respectively. These notable reductions in intralayer crossings favorably impact the overall loss budget and yield a more uniform ICL distribution. Such benefits are also applicable to higher node counts and help save hundreds of crossings compared to the lower bound for all-planar interconnects [Fig.~\ref{fig_scheme}(f), right].
Moreover, the proposed 3D routing structure exhibits inherent rotational symmetry: Layer 0 is centrally inverted, and Layer 1 is a $90^\circ$ rotation of Layer 0 [Fig.~\ref{fig_scheme}(a)-(b)]. This symmetry facilitates scalable design with higher waveguide density and more routing layers, as a fundamental unit can be replicated to generate the entire architecture.

\begin{figure*}[!t]
  \centering
  \includegraphics[width=0.9\linewidth]{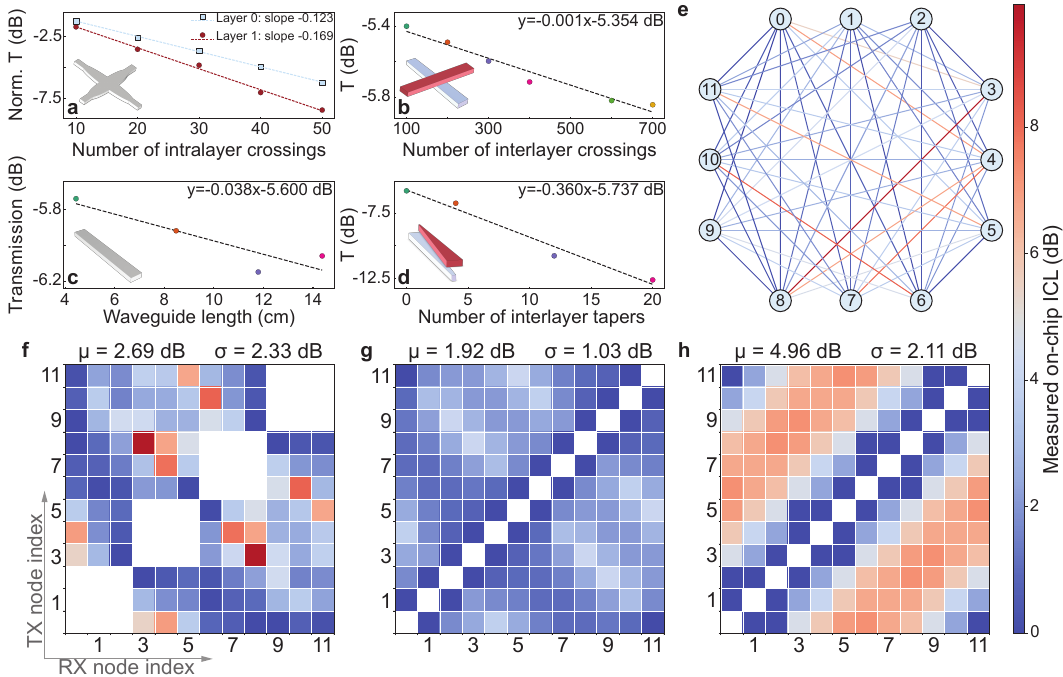}
  \caption{(a-d). C-band loss cutback measurements for: (a). intralayer crossing, (b). interlayer crossing, 
  (c). waveguide propagation loss, (d). interlayer taper.
  (e). Network graph of the measured on-chip ICL distribution. (f–h). On-chip ICL heatmaps with mean and standard deviation for: (f). measured results, 
  (g). predicted 3D interconnects with measured crossing loss and (h). predicted all-planar interconnects with measured crossing loss. Color bar represents on-chip ICL in dB and all losses are normalized by subtracting losses from edge couplers and external fiber paths.
  }
  \label{fig_measure}
\end{figure*}

The interposer is fabricated on a compact $7.4~\mathrm{mm}\times 7.4~\mathrm{mm}$ prototype die based on a dual-layer plasma enhanced chemical vapor deposition (PECVD) SiN platform. The nominal layer stack comprises two $400\,\mathrm{nm}$-thick SiN layers separated by a $1\,\text{\textmu m}$-thick SiO$_2$ spacer and capped by a $1.5\,\text{\textmu m}$-thick SiO$_2$ top cladding. Fig.~\ref{fig_dieshot}(a) shows the layout of the 3D SiN interposer, while Fig.~\ref{fig_dieshot}(b) and (c) present photos of the fabricated photonic interposer chip and a magnified view of the densely routed central region, respectively. Fig.~\ref{fig_dieshot}(d) summarizes the whole fabrication process flow. The fabrication begins with depositing $400\,\mathrm{nm}$ SiN using PECVD on a 4-inch Si wafer with a $3\,\text{\textmu m}$-thick $\text{SiO}_{\text{2}}$, followed by patterning of Layer~0. A tetraethoxysilane (TEOS)-based $\text{SiO}_{\text{2}}$ (around $1.5\,\text{\textmu m}$) is then deposited. A subsequent chemical mechanical planarization (CMP) step planarizes and thins the layer to form the $1\,\text{\textmu m}$-thick spacer. This sequence is repeated to form Layer~1. Throughout the fabrication process, stress-release patterns (SRPs) and high-temperature annealing steps are incorporated to relieve film stress~\cite{wuStressreleasedSi3N42020} and reduce propagation loss~\cite{jinDeuteratedSiliconDioxide2020a}. Finally, the wafer is diced into test-ready dies after deep silicon etch creating facets for fiber-to-chip coupling. Fig.~\ref{fig_dieshot}(e)-(g) display scanning electron microscopy (SEM) images of the key structures, including an intralayer crossing with elliptical multimode regions [Fig.~\ref{fig_dieshot}(e)] and a waveguide with a smooth sidewall [Fig.~\ref{fig_dieshot}(f)]. Fig.~\ref{fig_dieshot}(g) shows a cleaved cross section of the interlayer taper under a tilted viewing angle. The interlayer transition is implemented by a pair of complementary inverse tapers for adiabatic mode transfer. As the cross-sectional view is taken within the taper region, the different waveguide widths are expected. From the measurement in Fig.~\ref{fig_dieshot}(g), the fabricated SiN thicknesses are approximately $360\,\mathrm{nm}$ on Layer~0 and $270\,\mathrm{nm}$ on Layer~1, with an interlayer spacer thickness of $\sim 1.08\,\text{\textmu m}$ and a top cladding thickness of $\sim 0.43\,\text{\textmu m}$. Despite these deviations from the nominal values, the passive components remain functional with no qualitative change in operation. For instance, the measured intralayer crossing loss differs by only $0.046\,\mathrm{dB}$ per crossing between two layers [Fig.~\ref{fig_measure}(a)].

Fig.~\ref{fig_measure} summarizes the optical performance of the fabricated 3D SiN interposer.
Panels (a)-(d) report the cutback measurements of the key passive elements, showing $0.123\,\mathrm{dB}$ loss per intralayer crossing on Layer~0 and $0.169\,\mathrm{dB}$ on Layer~1, 
$0.001\,\mathrm{dB}$ loss of per interlayer crossing, $0.038\,\mathrm{dB/cm}$ waveguide loss and $0.360\,\mathrm{dB}$ loss per interlayer taper transition.
Fig.~\ref{fig_measure}(e) visualizes the measured on-chip ICL distribution in a network graph, providing an intuitive, topology-preserving view. Each line segment corresponds to an OIO-link and is colored according to the average loss of the waveguide pair. The same data are reorganized into an ICL heatmap indexed by the transmitter (TX) and receiver (RX) nodes [Fig.~\ref{fig_measure}(f)]. Cells corresponding to self-links (TX = RX) and OIO-links between the closely spaced nodes on the same die edge are left blank due to measurement limitations, and these entries are also excluded from all mean and standard deviation calculations.
Fig.~\ref{fig_measure}(g) and (h) present the predicted ICL heatmaps based on the measured losses of the passive components. Figure~\ref{fig_measure}(g) models the ICLs for the same optimized 3D dual-layer routing used in the prototype. It evaluates each waveguide according to Eq.~S4, using the layer-dependent intralayer crossing losses, the interlayer crossing loss and the interlayer taper loss [Fig.~\ref{fig_measure}(a), (b) and (d)]. Fig.~\ref{fig_measure}(h) illustrates the predicted baseline of all-planar interconnects with identical node placements [Fig.~\ref{fig_scheme}(c)], utilizing the average intralayer crossing loss of Layer~0 and Layer~1 [Fig.~\ref{fig_measure}(a)]. Waveguide propagation loss is neglected in both predictions because of its negligible contribution at the die scale.
These predictions therefore provide references for interpreting the prototype results, quantified by the mean $\text{\textmu}$ and standard deviation $\text{\textsigma}$ of the ICLs. 

The benefit of 3D routing over all-planar routing becomes clear: the measured prototype achieves $\text{\textmu}=2.69~\mathrm{dB}$, representing a remarkable 45.8\% reduction in the average loss per waveguide compared to the all-planar prediction $\text{\textmu} = 4.96~\mathrm{dB}$, although it exceeds the idealized 3D prediction $\text{\textmu}=1.92~\mathrm{dB}$ by $0.77~\mathrm{dB}$. In terms of uniformity, the measured spread $\text{\textsigma}=2.33~\mathrm{dB}$ is comparable to the baseline of all-planar interconnects $\text{\textsigma}=2.11~\mathrm{dB}$ but larger than the 3D prediction $\text{\textsigma}=1.03~\mathrm{dB}$.
The measured heatmap further reveals anomalously high losses in specific links (0–4, 3–8, 4–7, 4–8, 5–11, and 6–10), which may account for the larger mean and spread observed in the measurement compared to the idealized 3D prediction. These links match the high-ICL regions in the 3D prediction [Fig.~\ref{fig_measure}(g)], although the measured losses substantially exceed prediction. An explanation is that the edge couplers on the east side (ports 3–5) suffer more non-uniformity during fabrication and their actual coupling loss is noticeably higher than the calibrated value of edge couplers. The resulting excess coupling loss compounds with existing crossing losses, making these links appear especially lossy. This trend can be viewed in the color patterns in Fig.~\ref{fig_measure}(e) and (f), where east-side links have generally higher loss.
Other unmodeled factors, such as fluctuations in fiber-to-chip coupling and die-level non-uniformity of passive components (e.g., the thickness deviation from the design value shown in Fig.~\ref{fig_dieshot}(g)), can also contribute to the deviation between measurement and prediction. Collectively, these factors lead to the discrepancy between the measured performance and the idealized 3D prediction. While the prototype significantly outperforms the baseline of all-planar interconnects, further fabrication optimization is expected to bridge the gap toward the theoretical limit.

In conclusion, we propose a 3D photonic interposer architecture that overcomes the topological limits of all-planar routing. The experimental demonstration on the multilayer SiN platform confirms a significant reduction in the total crossing count (69.7\%) and the average loss per waveguide (45.8\%) over the all-planar baseline. With further fabrication optimization, the loss performance is expected to move closer to the model-based prediction of a $> 60\%$ reduction in average loss and a narrower loss distribution. The 3D SiN interposer can also be extended to other wavelengths, larger spatial scales, higher node counts, flexible node placements and additional routing layers to further reduce crossings. Our work thus provides a scalable solution for photonic interposers, addressing the escalating interconnect demands of future compute clusters and data centers.

\bibliography{sample}

\vspace{1 mm}
\noindent \textbf{Acknowledgments} We thank the funding support from the National Key R\&D Program of China (2024YFA1409300), the Research Grants Council of Hong Kong (C7143-25Y, N\_HKU774\_25, T46-705/23-R, STG3/E-704/23-N, STG3/E-104/25-N), the Innovation and Technology Commission of Hong Kong (GHP/230/22GD), the National Natural Science Foundation of China (6232290014), the Guangdong Provincial Quantum Science Strategic Initiative (GDZX2304004, GDZX2404002), and the Croucher Foundation.\\

\end{document}